\documentclass[12pt]{iopart}

\usepackage{iopams}
\usepackage{graphicx}% Include figure files
\usepackage{dcolumn}% Align table columns on decimal point
\usepackage{bm}% bold math
\usepackage{float}
\usepackage{color}
\newcommand{\bra}[1]{\langle#1|}

\newcommand{\ket}[1]{|#1\rangle}
\providecommand{\openone}{\leavevmode\hbox{\small1\kern-3.8pt\normalsize1}}
\begin{document}

\title{Spin-echo entanglement protection from random telegraph noise}

\author{R. Lo Franco$^{1,2}$, A. D'Arrigo$^3$, G. Falci$^3$,  G. Compagno$^2$ and E. Paladino$^3$}
\address{$^1$Centro Siciliano di Fisica Nucleare e Struttura della Materia (CSFNSM) 
Via Santa Sofia 64, 95123 Catania, Italy,\\
$^2$CNISM \& Dipartimento di Fisica, Universit\`a di Palermo, via Archirafi 36, 90123 Palermo, Italy,\\
$^3$Dipartimento di Fisica e Astronomia, Universit\`a di Catania, Via Santa Sofia 64, 95123 Catania, 
Italy \&  CNR-IMM UOS Catania (Universit\'a), Consiglio Nazionale delle Ricerche.
}
\ead{rosario.lofranco@unipa.it}

\begin{abstract}
We analyze local spin-echo procedures to protect entanglement between two
non-interacting qubits, each subject to pure-dephasing random telegraph noise. 
For superconducting qubits this simple model captures characteristic features of
the effect of bistable impurities coupled to the device. 
An analytic expression for the entanglement dynamics is reported. 
Peculiar features related to the non-Gaussian nature of the noise already observed in the
single qubit dynamics also occur in the entanglement dynamics for proper values of the ratio
$g=v/\gamma$, between the qubit-impurity coupling strength and the switching rate of the random
telegraph process, and of the separation between the pulses $\Delta t$.
We find that the echo procedure may delay the disappearance of entanglement, cancel the dynamical structure 
of entanglement revivals and dark periods, and induce peculiar plateau-like behaviors of the concurrence. 
\end{abstract}

%Uncomment for PACS numbers title message

% Keywords required only for MST, PB, PMB, PM, JOA, JOB?
%\vspace{2pc}
%\noindent{\it Keywords}: Article preparation, IOP journals
% Uncomment for Submitted to journal title message
%\submitto{\JPA}
% Comment out if separate title page not required

\maketitle

\section{\label{introduction}Introduction}

Precise control of multiple coupled quantum systems is a major goal towards the realization of 
quantum computation ~\cite{nielsenchuang,vandersypen2005RMP}. 
Multi-pulse sequences developed in the field of nuclear magnetic resonance (NMR) have 
recently been applied to mitigate noise in various qubits implementations ranging from atomic ensembles
to the solid state platform~\cite{echoSC,echo,bylander2011}.
When a multi-qubit system is considered, the main issue is to control the entanglement dynamics in order 
to  maintain a sufficient level of entanglement long-enough to efficiently perform two-qubit operations 
and entanglement storage. On a more fundamental level, the issue is the possibility to prevent
entanglement disappearance, i. e. the phenomenon of entanglement sudden death (ESD)~\cite{yu2009Science}.

These problems, shared to a certain extent by all qubit implementations, are particularly
severe for solid state qubits which suffer from material inherent noise sources.
In particular, superconducting qubits are sensitive to fluctuating impurities located in the insulating
materials surrounding superconducting islands, or inside the 
junctions. Often impurities originate 
random telegraph (RT) fluctuations of islands polarizations~\cite{Zorin1996,Nakamura1999} or 
of magnetic fluxes in SQUID geometries~\cite{Yoshihara2006}. 
The noise spectrum of the corresponding variables is a Lorentzian centered at zero frequency.
An ensemble of impurities may originate the $1/f$ low frequency behavior of the noise spectrum observed in 
several nanodevices~\cite{weissman1988RMP}.
Large amplitude noise at low frequencies gives rise to dephasing, which is due to the randomization of 
the dynamic phase difference between superpositions of the qubit computational states. This phenomenon,
as a difference with irreversible energy relaxation, is in principle reversible and can be
refocused dynamically through the application of coherent control-pulse methods \cite{vandersypen2005RMP}.
The possibility to extend to superconducting nanocircuits decoupling techniques developed in NMR 
has been demonstrated in different labs~\cite{echoSC,bylander2011,nakamura2002}.

The simplest decoupling procedure is the spin-echo sequence \cite{Hahn1950}. It was successfully 
implemented in a charge qubit already in 2002~\cite{nakamura2002}, unambiguously proving that 
low-frequency energy-level 
fluctuations due to $1/f$ charge noise were the main source of dephasing.
Since two-qubit correlations are more fragile than single qubit coherence, it is natural to ask
whether a generalization of the spin-echo technique can be exploited to 
maintain entanglement between two solid-sate qubits in the presence of low frequency noise.
This analysis falls within the rapidly developing research area aiming to 
extend pulse-based dynamical decoupling procedures~\cite{viola1999PRL}
to the entanglement dynamics~\cite{DD-entanglement}. 
Recently, dynamical decoupling techniques to mitigate noise and enhance the lifetime of entangled 
state that is formed in a superconducting flux qubit coupled to a microscopic two-level
system have been implemented~\cite{gustavsson2012}. 

Here we address this issue considering a simple model system which captures peculiar features of the effects
of impurities on superconducting qubits: each qubit is longitudinally coupled (pure dephasing), 
with coupling strength $v$, to a bistable fluctuator randomly switching between the two 
states with rate $\gamma$~\cite{paladino2002PRL,galperin2003,galperin2006PRL}.  
The non-Gaussian nature of the RT process clearly manifests itself
when the Markovian approximation does not apply, that is when the fluctuator is ``slow'' enough.
More precisely, the Markovian approximation breaks down when the ratio between the
coupling strength and the switching rate, $g=v/\gamma$, is sufficiently larger than one. When this
condition is met, the single qubit dynamics displays non-exponential decay, beatings, and
dependence on the system initial conditions~\cite{paladino2002PRL,paladino2003,bergli2009}. 
In  Ref.~\cite{galperin2006PRL} the effect of an echo procedure on a superconducting qubit affected by 
a RT noise source at pure dephasing has been studied. In the non-Gaussian regime the echo signal as a
function of time shows  a series of plateaus whose position and eights depend separately on $v$ and
$\gamma$. In general, when $g \geq 1$ the echo signal differs significantly from the prediction based
on a Gaussian approximation of the stochastic process.

In this paper we investigate the possibility to preserve entanglement between two
Josephson qubits each subject to a RT fluctuator at pure dephasing, by applying simultaneous
echo pulses to the two qubits.
We shall show that the echo procedure may delay the disappearance of 
entanglement and cancel the dynamical structure of entanglement revivals followed by dark periods, 
exhibiting also a peculiar plateau-like behavior.

\section{Hamiltonian model\label{models}}
We consider two noninteracting superconducting qubits, $A$ and $B$, each
longitudinally coupled to a bistable impurity~\cite{lofranco2012PhysScripta}.
 We are interested to the regime where
the impurity splitting is smaller than the temperature, so that the impurity behaves as a
classical RT fluctuator \cite{paladino2002PRL,bergli2009,falci2005PRL}. 

The total Hamiltonian of our system is given by
$H_\mathrm{tot}=H_A+H_B$, where the Hamiltonian of each qubit affected by the RT impurity, $H_\alpha$ ($\alpha=A,B$), 
is  
\begin{equation}\label{HamiltonianQ}
H_\alpha=H_{Q, \alpha}+\mathcal{V}_\alpha(t),\quad
H_{Q, \alpha}=-(\Omega_\alpha/2)\sigma_{z, \alpha} - (v_\alpha \xi_\alpha(t)/2)\sigma_{z, \alpha},
\end{equation}
where $\xi_\alpha(t)$  instantly switches between $0$ and $1$ at random times with a
switching rate $\gamma_\alpha$
and  $v_\alpha$ is the coupling constant of qubit-$\alpha$ with a nearby impurity. 
The power spectrum of the unperturbed equilibrium fluctuations of $\xi_\alpha(t)$ is a Lorentzian,  $s_\alpha(\omega)=v_\alpha^2\gamma_\alpha/[2(\gamma_\alpha^2+\omega^2)]$,
and $\mathcal{V}_\alpha(t)$ denotes an external control field.
 
Due to the longitudinal coupling, the diagonal elements of the reduced density matrix of each qubit
(populations) are constant, whereas the off-diagonal elements (coherences) decay.
The noisy dynamics has been studied in \cite{paladino2002PRL,galperin2003,bergli2009}, where
the single-qubit coherence, $q_{0, \alpha}(t)$, has been found in analytic form
\begin{equation}\label{singlequbitcoherence}
\frac{q_{0, \alpha}(t)}{q_{0, \alpha}(0)}= e^{-i\Omega t} \, e^{-i\frac{v_\alpha t}{2}}
[A_\alpha e^{-\frac{\gamma_\alpha(1-\mu_\alpha)t}{2}}
+(1-A_\alpha)e^{-\frac{\gamma_\alpha(1+\mu_\alpha)t}{2}}],
\end{equation}
where $A_\alpha=\frac{1}{2\mu_\alpha}(1+\mu_\alpha - ig_\alpha \delta p_{0, \alpha})$, 
$\mu_\alpha=\sqrt{1-g_\alpha^2}$ and $\delta p_{0, \alpha}$ is the initial population difference of impurity-$\alpha$'s 
states.
Based on Eq. (\ref{singlequbitcoherence}), two regimes can be identified,
depending on the value of the parameter $g_\alpha=v_\alpha/\gamma_\alpha$.
When $g_\alpha \ll 1$ the impurity behaves as a weakly coupled and short-time correlated noise source
affecting the qubit. The coherence decays exponentially with a decoherence rate given by the
golden rule rate, $\propto v_\alpha^2/\gamma_\alpha$. In this regime the discrete process can be approximated 
as a Gaussian stochastic process completely characterized by the noise spectrum $s_\alpha(\omega)$. 
It is common to refer to this situation as ``weakly coupled'' impurity~\cite{paladino2002PRL}.
For $g_\alpha \geq 1$ instead the discrete nature of the process shows up in the qubit time evolution
which shows beatings at frequencies $\Omega_\alpha \pm v_\alpha$ and a long time decay with rate
$\gamma_\alpha$. 
In this ``strong coupling'' regime in fact the fluctuator splits the qubit's levels and the qubit
just experiences rare hops between these states with hopping rate $\gamma_\alpha$. 
 
Control is operated as in Ref.~\cite{viola98}, the external field $\mathcal{V}_\alpha(t)$ being a
sequence of  $\pi$-pulses about $\hat x$. We consider a modified spin-echo protocol consisting of 
two consecutive $\pi$-pulses separated by an interval $\Delta t$. 
The external pulses are short enough for both relaxation and spectral diffusion during each of 
the pulses to be neglected, thus the pulse  evolution operator is 
${\mathcal S}_{P, \alpha} \approx \exp{(i \pi/2 \sigma_{x \alpha})}= i \sigma_{x \alpha}$.
The evolution between the pulses reads ${\mathcal S}_{\alpha} = 
{\mathcal T}\{\int_0^{\Delta t} \exp[i {\mathcal H}_\alpha(t^\prime)] d t^\prime \}$. 
The basic idea behind the echo procedure is that
the sequence of two $\pi$ pulses about $\hat x$ reverses the sign of the qubit-fluctuator interaction during the
two time intervals $\Delta t$. This is clearly seen considering the short $\Delta t$ limit of 
${\mathcal S}_{\alpha} \propto i \sigma_{x \alpha} \Delta t $ and the composition of Pauli matrices
$\sigma_x \sigma_z \sigma_x = - \sigma_z$.
As a result, the effect of fluctuations slower than $\Delta t$ is cancelled at time
$2 \Delta t$, the residual signal decay being due to faster noise components. 
The echo signal for a qubit subject to a RT fluctuator at pure dephasing has been evaluated both in 
Ref.~\cite{falci2004PRA}, starting from a quantum description of the impurity, and in Ref.~\cite{galperin2006PRL}.
Here we report the final form of the single qubit coherence $q_{\mathrm{e}, \alpha}(t)$
\begin{equation}\label{coherenceecho}
\frac{q_{\mathrm{e}, \alpha}(2 \Delta t)}{q_{\mathrm{e}, \alpha}(0)}=
\frac{e^{-\gamma_\alpha\Delta t}}{\mu_\alpha^2}\left[\frac{1+\mu_\alpha}{2}e^{\mu_\alpha \gamma_\alpha\Delta t}+
\frac{1-\mu_\alpha}{2} e^{-\mu_\alpha\gamma_\alpha\Delta t}-(1-\mu_\alpha^2)\right] \, .
\end{equation}
Notice that for $\gamma_\alpha \Delta t\rightarrow 0$ one obtains $q_{\mathrm{e}, \alpha}(2\Delta t)\rightarrow 1$, 
as expected when the pulse separation $\Delta t$ is much smaller than the noise correlation time $1/\gamma_\alpha$.
In the following we shall use the above analytic expressions of single-qubit coherences to study the 
time-behavior of entanglement.

\section{Dynamics of entanglement and entanglement echo}
Based on the single qubit coherence reported in the previous Section, we investigate the time behavior
of entanglement between the two qubits. We consider the simple situation where the two qubits $A$ and $B$, initially
prepared in an entangled state, evolve independently each subject to a RT process, as described by the Hamiltonian
(\ref{HamiltonianQ}). We suppose that each qubit is subject to an echo sequence, as described above.
For the sake of simplicity, we suppose that the pulses are applied simultaneously to the two qubits and
evaluate the concurrence at time $2 \Delta t$. The entanglement echo is compared with dynamics of entanglement 
when no pulses are applied.
To this end, we need the evolved two-qubit density matrix. Since the two qubits are non-interacting,
their density matrix can be obtained following a standard procedure based on the knowledge of single-qubit 
dynamics~\cite{bellomo2007PRL,lofranco2012review}. This procedure also holds when the qubits are subject to 
independent external control fields, for the case of our interest, to two echo procedures.  

\subsection{Concurrence for non-interacting qubits}
We take as initial states the class of extended Werner-like (EWL) 
states \cite{bellomo2010PLA}
\begin{equation}\label{EWLstates}
   \rho_1=r \ket{1_{a}}\bra{1_{a}}+\frac{1-r}{4}\openone_4,\quad
    \rho_2=r \ket{2_{a}}\bra{2_{a}}+\frac{1-r}{4}\openone_4,
\end{equation}
whose pure parts are the one-excitation and two-excitation Bell-like states $\ket{1_{a}}=a\ket{01}+b\ket{10}$, 
$\ket{2_{a}}=a\ket{00}+b\ket{11}$, where $|a|^2+|b|^2=1$. The purity parameter $r$ quantifies the  purity of 
the state, given by $P=(1+3r^2)/4$. The density matrix of EWL states, in the computational basis 
$\mathcal{B}=\{\ket{0}\equiv\ket{00},\ket{1}\equiv\ket{01}, \ket{2}\equiv\ket{10}, \ket{3}\equiv\ket{11}\}$ 
has an X form~\cite{bellomo2008nonlocal} and this structure is maintained at $t>0$ during the pure-dephasing dynamics we are 
considering here. 
The initial entanglement is equal for both the EWL states of Eq.~(\ref{EWLstates}), 
$C_{\rho_1}(0)=C_{\rho_2}(0)=2\mathrm{max}\{0,(|ab|+1/4)r-1/4\}$, where $C$ is the concurrence \cite{wootters1998PRL}. 
Initial states are thus entangled for $r>r^\ast=(1+4|ab|)^{-1}$.

The EWL states of Eq.~(\ref{EWLstates}) evolve with fixed diagonal elements and time-dependent anti-diagonal 
elements given by, respectively, $\rho_{12}(t)=\rho_{12}(0)|q_A(t)q_B(t)|$ for the initial state $\rho_1$ and 
$\rho_{03}(t)=\rho_{03}(0)|q_A(t)q_B(t)|$ for $\rho_2$, where $q_\alpha(t)$ are given either by 
Eq.~(\ref{singlequbitcoherence})  or by Eq.~(\ref{coherenceecho}), depending on the qubit evolution.
The general expressions of the concurrences at time $t$ for 
the two initial states of Eq.~(\ref{EWLstates}) are easily obtained as 
$C_{\rho_1}(t)=2\mathrm{max}\{0,|\rho_{12}(t)|-\sqrt{\rho_{00}(0)\rho_{33}(0)}\}$ and 
$C_{\rho_2}(t)=2\mathrm{max}\{0,|\rho_{03}(t)|-\sqrt{\rho_{11}(0)\rho_{22}(0)}\}$ \cite{lofranco2012PhysScripta}. 
These expressions coincide for the initial states $\rho_1$ and $\rho_2$,
$C_{\rho_1}(t)=C_{\rho_1}(t)=C(t)$ where 
\begin{equation}\label{Kt}
C(t)=2\mathrm{max}\{0,r|a|\sqrt{1-|a|^2}|q_{A}(t) q_{B}(t)|-(1-r)/4\}.
\end{equation}
For initial pure states, $r=1$, it is readily seen from Eq.~(\ref{Kt}) that $C(t)\propto |q_A(t)q_B(t)|$ 
so that entanglement qualitatively behaves like the single-qubit coherence. 
In a more realistic case, the initial state is not pure. 
Here we consider a realistic degree of purity in superconducting systems. We refer to the experiment \cite{schoelkopf2009Nature}
where entangled states of two Josephson qubits with purity $\approx 0.87$ and fidelity to ideal Bell states 
$\approx 0.90$ have been generated by two-qubit interaction mediated by a cavity bus in a circuit quantum electrodynamics 
architecture and tunable in strength by two orders of magnitude on nanosecond timescales. 
These states may be approximately described as EWL states with $r=r_\mathrm{exp}\approx0.91$ and $|a|=1/\sqrt{2}$, giving
a concurrence  $C=0.865$~\cite{palermocatania2010PRA}. 
In the following analysis we shall assume the qubits are prepared in an initial entangled state having these
characteristics.

\subsection{Entanglement echo}
The advantage of the entanglement echo procedure and the peculiarities resulting from the effect
of discrete stochastic processes affecting the two qubits are pointed out by comparing with
the entanglement evolution at pure dephasing in the presence of RT fluctuations which has been
studied in Ref.~\cite{lofranco2012PhysScripta}. 
Under these conditions the concurrence is given by Eq.~(\ref{Kt}) with $q_{0,\alpha}(t)$ as in 
Eq.~(\ref{singlequbitcoherence}). In the following analysis we put $ \delta p_{0, \alpha}=0$.

\begin{figure}
\begin{center}
{\includegraphics[width=0.35\textwidth]{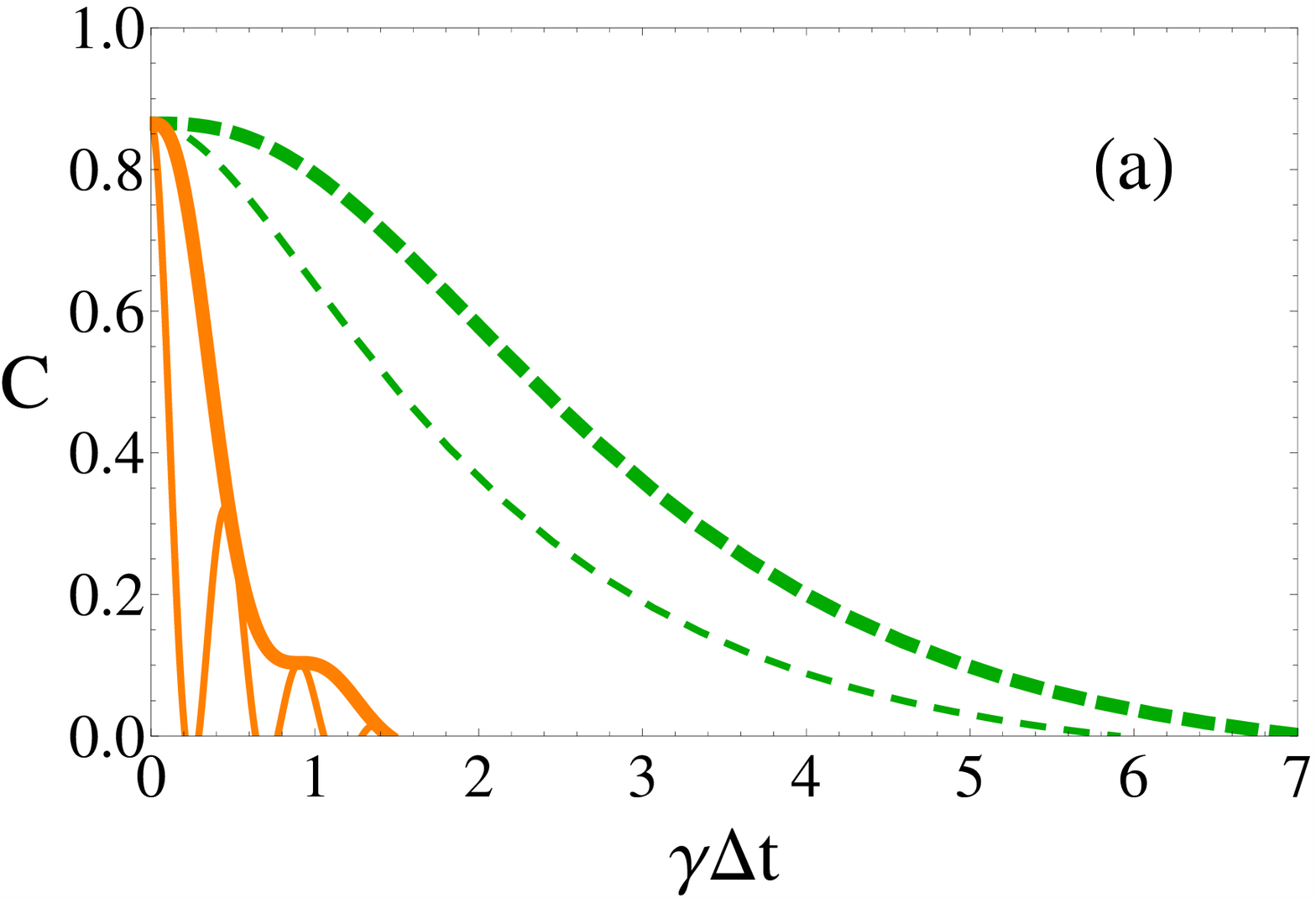}\hspace{0.5 cm}
\includegraphics[width=0.35\textwidth]{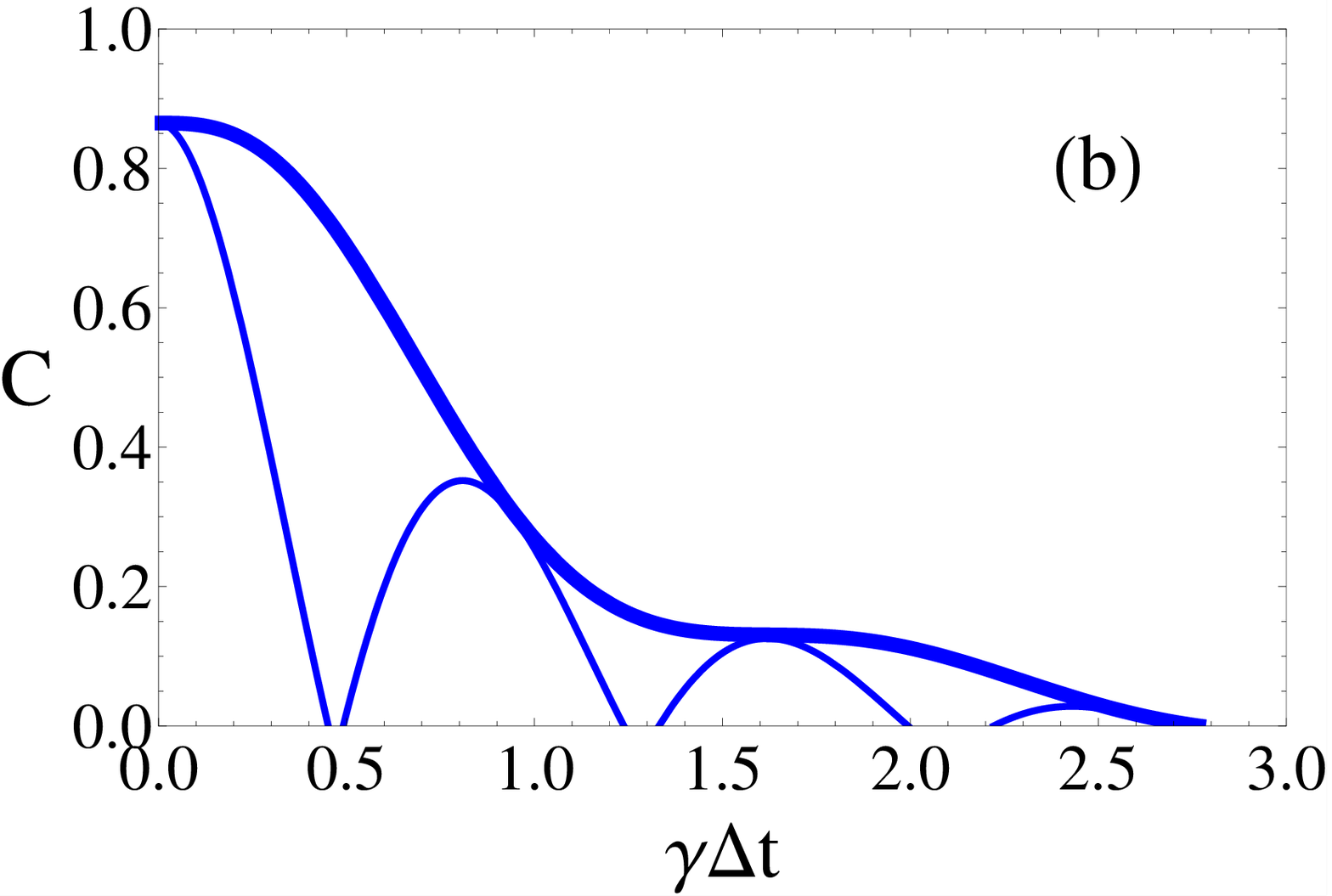}}
\caption{\label{fig:Ctime}\footnotesize (Color online) 
Concurrence $C(2\Delta t)$ as a function of $\gamma\Delta t$.
In \textbf{(a)} $g_\alpha=g$, with $g_1=0.7$ (dashed green lines) and $g_2=7$ (continuous orange lines), 
for the cases without echo (thin lines) and with echo (thick lines). For $g_2=7$ note the plateaus at 
$\gamma \Delta t = 2 \pi/g_2 \approx 0.9$; In \textbf{(b)} $g_A= 0.3$ and $g_B=4$ 
for the cases without echo (thin line) and with echo (thick line).}
\end{center}
\end{figure}

As a difference with the single qubit evolution, where the threshold separating Gaussian and non-Gaussian
behavior is at $g \sim 1$, the entanglement dynamics displays qualitative different features
at a threshold value depending on the characteristics of the initial entangled state, 
$\bar{g}(r,a)$. For initial mixed states, $r<1$, the time-behavior of entanglement qualitatively 
changes in correspondence of a threshold value $\bar{g}(r,a)>1$~\cite{lofranco2012PhysScripta}. 
There is always ESD with exponential decay for $g\leq\bar{g}$ and a complete disappearance with revivals for 
$g>\bar{g}$ (thin curves of Fig.~\ref{fig:Ctime}(a)). 
For the considered EWL initial states the threshold is $\bar{g}=2.3$. We remark  that 
the entanglement revivals here occur in a classical environment unable of any back-action. 
The interpretation of this effect differs from the 
explanation valid in the presence of quantum environments \cite{lopez2008PRL} and 
it is an open issue \cite{lofranco2012PRA,darrigo2012arxiv}.

We now act on each qubit with two simultaneous echo procedures.
For sake of simplicity, we suppose that the two fluctuators have equal
switching rates, $\gamma_\alpha \equiv \gamma$, but  different couplings
to the corresponding qubit, in order to address the regime $g_A \neq g_B$.
The concurrence $C(t=2\Delta t)$ is found in this case by Eq.~(\ref{Kt}) with 
$q_{\mathrm{e}, \alpha}(t)$ of Eq.~(\ref{coherenceecho}).

To start with, we consider the case $g_\alpha \equiv g$ and evaluate $C$ for two values
$g_1 < \bar{g}$ and $g_2 > \bar{g}$.
In Fig.~\ref{fig:Ctime} we plot the concurrences with and without echo as a function of 
$\gamma\Delta t$. To point out clearly the qualitative  effects we take very different 
values, $g_1=0.7$ and $g_2= 7$. 
We observe that the echo procedure leads to  a larger degree of entanglement with respect to the free evolution 
for both values of $g$ considered. 
In particular, for $g_1\leq\bar{g}$ the ESD time is delayed (dashed lines in Fig.~\ref{fig:Ctime}(a)). 
For $g_2>\bar{g}$ the dynamical structure of entanglement revivals and dark periods is cancelled out by the echo 
procedure. Moreover, plateau-like features appear in the time behavior of entanglement (continuous lines in 
Fig.~\ref{fig:Ctime}(a)). This is a typical non-Gaussian behavior linked to the plateaus of the single-qubit coherence
observed in this range of $g$ predicted in Ref.~\cite{galperin2006PRL}, where it is also suggested that
a similar feature occurs in the experiment of Ref.~\cite{nakamura2002}.  
In fact, when $g\gg 1,\sqrt{1/(\gamma\Delta t)}$ the qubit coherence $q_{\mathrm{e}, \alpha}(t)$ of Eq.~(\ref{coherenceecho})
acquires the simple form \cite{galperin2006PRL}
\begin{equation}\label{coherencelargeg}
q_{\mathrm{e}, \alpha}(t)\approx e^{-\gamma_\alpha\Delta t}[1+\sin(g_\alpha \gamma_\alpha \Delta t)/g_\alpha],
\end{equation}
which in turns gives an approximate expression of concurrence $C(t)$ by Eq.~(\ref{Kt}). The plateau-like behavior 
in the concurrence $C(t=2\Delta t)$ occurs when $\mathrm{d}C/\mathrm{d}(\Delta t)\approx 0$ that is, 
being $\mathrm{d}C/\mathrm{d}(\Delta t)\propto\mathrm{d}q_\mathrm{e}^2/\mathrm{d}(\Delta t)=2q_\mathrm{e}
[\mathrm{d}q_\mathrm{e}/\mathrm{d}(\Delta t)]$, 
when $\mathrm{d}q_\mathrm{e}/\mathrm{d}(\Delta t)\approx0$. The latter corresponds to the plateau-like 
behavior of the single-qubit coherence predicted in Ref.~\cite{galperin2006PRL}. 
Using $q_{\mathrm{e}, \alpha}$ of Eq.~(\ref{coherencelargeg}) one gets 
$g_\alpha \gamma_\alpha \Delta t_\mathrm{plateaus} \approx 2\pi k$ ($\gamma_\alpha t_\mathrm{plateaus}\approx 4k\pi/g_\alpha$). 
From Eqs.~(\ref{Kt}) and (\ref{coherencelargeg}), we also find that the values taken by the concurrence in correspondence
to the plateaus (with $g_\alpha\equiv g$) are 
$C(t_\mathrm{plateau})\approx 2r|a|\sqrt{1-|a|^2}e^{-4 k\pi/g}-(1-r)/4$.
Thus when $g \gg 1$, the entanglement at times $t$ close to $4k\pi/v$ is almost insensitive 
to small variations of $\Delta t$. 
This simple effect of entanglement echo is a new feature peculiar of entanglement under pure dephasing due to a 
discrete noise process. 

We now consider the effect of echo on the entanglement dynamics when one fluctuator is characterized by
a $g_\alpha$ below and the other above $\bar g$. In Fig.~\ref{fig:Ctime}(b) we plot $C$, without and with echo, 
as a function of $\gamma \Delta t$ for $g_A= 0.3$ and $g_B = 4$. 
The echo produces the same qualitative plateau-like behavior observed when $g_\alpha > \bar g$.
The effect however is less pronounced since the plateaus originate only from the echo on the qubit
affected by the strongly coupled fluctuator, the echo on the weakly coupled fluctuator merely slowing down the
coherence decay thus modulating the concurrence amplitude at the plateaus.
   
\begin{figure}
\begin{center}
{\includegraphics[width=0.40\textwidth]{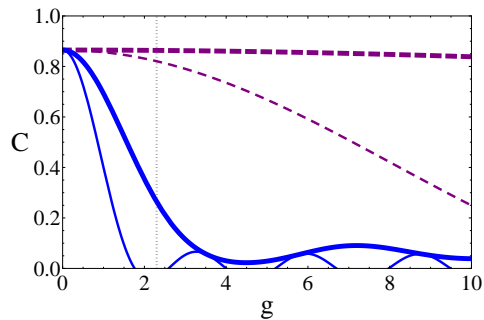}}
\caption{\label{fig:Cg}\footnotesize (Color online) Concurrence $C$ as a function of $g$ for two values of $\gamma\Delta t$ 
equal to $0.1$ (purple dashed lines) and $1.1$ (blue solid lines) for the cases without echo (thin lines) and with echo (thick lines). The dotted vertical line corresponds to $\bar{g}=2.3$.}
\end{center}
\end{figure}

Finally we investigate the dependence of the concurrence on $g \equiv g_\alpha$ at fixed values of the echo pulse
interval $\gamma \Delta t$, shown in Fig.~\ref{fig:Cg} for $\gamma \Delta t= 0.1, 1.1$.
As expected, when $\gamma\Delta t\ll 1$ the echo procedure is very effective in suppressing the noise, even 
for relatively large values of $g$ (dashed lines in Fig.~\ref{fig:Cg}).
For intermediate values of $\gamma\Delta t \sim 1$ instead  we observe a non-monotonous behavior of $C$.
The oscillatory behavior of $C$ at fixed $\gamma \Delta t$
when $g \gg 1$ is simply understood considering the large $g$ limit of the coherence Eq. (\ref{singlequbitcoherence}),
$|q_{0, \alpha}| \approx \exp{(- \gamma_\alpha t/2)} \cos{(g_\alpha \gamma_\alpha t/2)}$.
Interestingly, the echo preserves entanglement even when $g>\bar{g}$, when it would vanish in the absence of  
the echo procedure (continuous lines in Fig.~\ref{fig:Cg}). 
However, the quantitative value of preserved  entanglement might not be sufficient for efficient realization of
quantum error correction tasks.
Finally, for large values of the pulse intervals, $\gamma\Delta t\gg1$, the echo procedure is not able to efficiently 
reduce the detrimental effects of  noise for any $g$.

\section{Conclusions \label{conclusion}}
In this paper we have investigated the effect of the simultaneous application of a modified version of
the spin-echo protocol to two noninteracting qubits each subject to pure-dephasing RT noise.
This simple system is inspired to superconducting qubits, where selected impurities producing RT noise are frequently observed 
and techniques from NMR have been recently implemented, showing the possibility
to limit defocusing due to low-frequency noise and to strongly coupled 
impurities~\cite{echoSC,bylander2011,nakamura2002}.

Here we presented an analytic expression for the concurrence in the presence of
simultaneous echo protocols on the two qubits for a general class of initial entangled states.
The main result is that the echoes either delay the ESD time or cancel the dynamical structure 
of entanglement revivals followed by dark periods, depending on the 
qubit-fluctuator coupling strength $g_\alpha$. 
In particular, when for at least one qubit $g_\alpha > \bar g$, the entanglement exhibits a novel dynamical 
structure consisting of  plateaus occurring at selected values of pulse separation $\Delta t$. 
This effect is entirely due to the non-Gaussian nature of the noise. 
For identical qubits and impurities the plateaus times are at 
$v_\alpha \Delta t_\mathrm{plateaus} \approx 2\pi k$.
In general, when the pulse lengths are very short, $1/\Delta t\gg\gamma_\alpha$, the two echoes considerably 
slow down  the entanglement decay up to relatively large values of $g_\alpha \gtrsim \bar g$. Interestingly, for 
intermediate pulse lengths ($\gamma\Delta t\sim 1$) entanglement can be at least partly preserved even 
for $g>\bar{g}$ when it vanishes in the absence of the echo procedure.

This very simple analysis aimed at pointing out relevant qualitative effects, starting from a physically
relevant model. Relevant issues like the feasibility of simultaneous pulses, the effect of timing imperfections,
the interplay with additional low-frequency components leading to $1/f$ noise have not been addressed. 
At the single qubit level the echo protocol can limit defocusing due to $1/f$ noise~\cite{echoSC,bylander2011,nakamura2002}. 
An analysis of the effect of $1/f$ noise on the entanglement both of uncoupled~\cite{palermocatania2010PRA} and 
of coupled qubits \cite{coupledCT} has been recently performed.  
This preliminary analysis suggests that the echo protocol and dynamical decoupling extensions 
may be conveniently exploited to protect entanglement between solid state qubit, and possibly
other kinds of correlations like those quantified by the quantum discord \cite{modi2012review},
against random telegraph and $1/f$ noise~\cite{lofrancoinprogress}.  
%Another development is then the analysis 
%of these pulse-based procedures to protect other kinds of correlations, for example those quantified by 
%the quantum discord \cite{modi2012review}.   

\section*{Acknowledgments}
Partially supported by EU through Grant No. PITN-GA-2009-234970 and by the joint Italian-Japanese Laboratory on
``Quantum Technologies: Information, Communication and Computation'' of the 
Italian Ministry of Foreign Affairs.

\section*{References}

\providecommand{\newblock}{}


\begin{thebibliography}{10}
\expandafter\ifx\csname url\endcsname\relax
  \def\url#1{{\tt #1}}\fi
\expandafter\ifx\csname urlprefix\endcsname\relax\def\urlprefix{URL }\fi
\providecommand{\eprint}[2][]{\url{#2}}
% Bibliography created with iopart-num v2.1
% /biblio/bibtex/contrib/iopart-num

\bibitem{nielsenchuang}
Nielsen M~A and Chuang I~L 2000 {\em Quantum Computation and Quantum
  Information\/} (Cambridge University Press).
  
\bibitem{vandersypen2005RMP}
Vandersypen L~M~K and Chuang I~L 2005 {\em Rev. Mod. Phys.\/} {\bf 76} 1037

\bibitem{echoSC}
Chiorescu I. {\em et al.} 2003
% Y. Nakamura, C. J. P. M. Harmans, and J. E. Mooij, 
{\em Science} {\bf 299} 1869;
Ithier G {\em et al.} 2005 {\em Phys. Rev. B} {\bf 72} 134519;
 Biercuk M J {\em et~al.\/} 2009 {\em Nature \/} {\bf 458} 996;
  Biercuk M J {\em et~al.\/} 2009 {\em Phys. Rev. A \/} {\bf 79} 062324.

\bibitem{echo} 
Petta J R %A. C. Johnson, J. M. Taylor, E. A. Laird,
%A. Yacoby, M. D. Lukin, C. M. Marcus, M. P. Hanson, and A. C. Gossard, 
{\em et al.} 2005 {\em Science} {\bf 309} 2180;
de Lange J {\em et~al.\/} 2012 {\em Sci. Rep.\/} {\bf 328} 1.
 
\bibitem{bylander2011}
 Bylander G {\em et~al.\/} 2011 {\em Nat. Phys.\/} {\bf 7} 565

\bibitem{yu2009Science}
Yu T and Eberly J~H 2009 {\em Science\/} {\bf 323} 598


\bibitem{Zorin1996} Zorin A B {\em et al.} 1996 {\em Phys. Rev. B} {\bf 53} 13682
\bibitem{Nakamura1999} Nakamura Y {\em et al.} 1999 {\em Nature} {\bf 398}, 786

\bibitem{Yoshihara2006} 
Yoshihara F {\em et al.} 2006 {\em Phys. Rev. B}
{\bf 81} 132502;
Kakuyanagi K {\em et al.} 2007 {\em Phys. Rev. Lett.} {\bf 98} 047004;
Bialczak R C 2007 {\em Phys. Rev. Lett.} {\bf 99} 187006;
Lanting T {\em et al.} 2009 {\em Phys. Rev. B} {\bf 79} 060509(R)


\bibitem{weissman1988RMP}
Weissman M~B 1988 {\em Rev. Mod. Phys.\/} {\bf 60} 537

\bibitem{nakamura2002}
Nakamura Y {\em et~al.\/} 2002 {\em Phys. Rev. Lett.\/} {\bf 88} 047901


\bibitem{Hahn1950}
Hahn E 1988 1950 {\em Phys. Rev.\/} {\bf 80} 580


\bibitem{viola1999PRL}
Viola L, Knill E and Lloyd S 1999 {\em Phys. Rev. Lett.\/} {\bf 82} 2417

\bibitem{DD-entanglement}
Mukhtar M, Soh W~T, Saw T~B and Gong J 2010 {\em Phys. Rev. A\/} {\bf 82} 052338;
Agarwal G~S 2010 {\em Phys. Scr.\/} {\bf 82} 038103;
Wang Z~Y and Liu R~B 2011 {\em Phys. Rev. A\/} {\bf 83} 022306;
Pand Y, Xi Z-R and Gong J 2011 {\em J. Phys. B: At. Mol. Opt. Phys.\/} {\bf 44}
175501.

\bibitem{gustavsson2012}
Gustavsson S {\em et~al.\/} 2012 arXiv:1204.6377

\bibitem{paladino2002PRL}
Paladino E, Faoro L, Falci G and Fazio R 2002 {\em Phys. Rev. Lett.\/} {\bf 88}
  228304

\bibitem{galperin2003} 
Galperin Y~M, Altshuler B~L  and Shantsev D~V 2003 cond-mat/0312490v1.

\bibitem{galperin2006PRL}
Galperin Y~M, Altshuler B~L, Bergli J and Shantsev D~V 2006 {\em Phys. Rev.
  Lett.\/} {\bf 96} 097009


\bibitem{paladino2003}
Paladino E {\em et~al.\/} 2003 {\em Adv. Sol. St. Ph. \/} {\bf 43}
747

\bibitem{bergli2009}
Bergli J, {\em et~al.\/} 2009 {\em New J. Phys.\/} {\bf 11} 025002

\bibitem{lofranco2012PhysScripta}
{Lo Franco} R, {D'Arrigo} A, Falci G, Compagno G and Paladino E 2012 {\em Phys.
  Scr.\/} {\bf T147} 014019  

\bibitem{falci2005PRL}
Falci G, {D'Arrigo} A, Mastellone A and Paladino E 2005 {\em Phys. Rev.
  Lett.\/} {\bf 94} 167002


\bibitem{falci2004PRA}
Falci G, D'Arrigo A, Mastellone A and Paladino E 2004 {\em Phys. Rev. A\/} {\bf
  70} 040101(R)

\bibitem{viola98}
Viola L, Knill E and Lloyd S 1998 {\em Phys. Rev. A \/} {\bf 58} 2733.

\bibitem{bellomo2007PRL}
Bellomo B, {Lo Franco} R and Compagno G 2007 {\em Phys. Rev. Lett.\/} {\bf 99}
  160502

\bibitem{lofranco2012review}
{Lo Franco} R, Bellomo B, Maniscalco S and Compagno G arXiv:1205.6419

%\bibitem{Yu2004}
%Yu T and Eberly J~H 2004 {\em Phys. Rev. Lett. \/} {\bf 93} 140404.


\bibitem{bellomo2010PLA}
Bellomo B, {Lo Franco} R and Compagno G 2010 {\em Phys. Lett. A\/} {\bf 374}
  3007

\bibitem{bellomo2008nonlocal}
Bellomo B, {Lo Franco} R and Compagno G 2009 {\em Adv. Sci. Lett.\/} {\bf 2}
  459

\bibitem{wootters1998PRL}
Wootters W~K 1998 {\em Phys. Rev. Lett.\/} {\bf 80} 2245

\bibitem{schoelkopf2009Nature}
DiCarlo L {\em et al.} 2009 {\em Nature\/} {\bf 460} 240
%Chow J~M, Gambetta J~M, Bishop L~S, Johnson B~R, Schuster D~I, Majer J, Blais A, Frunzio L, Girvin S~M and Schoelkopf R

\bibitem{palermocatania2010PRA}
Bellomo B {\em et al.} 2010 {\em Phys. Rev. A\/} {\bf 81} 062309
%, Compagno G, D'Arrigo A, Falci G, {Lo Franco} R and Paladino E


\bibitem{lopez2008PRL}
L{\'{o}}pez C~E, Romero G, Lastra F, Solano E and Retamal J~C 2008 {\em Phys.
  Rev. Lett.\/} {\bf 101} 080503

\bibitem{lofranco2012PRA}
{Lo Franco} R, Bellomo B, Andersson E and Compagno G 2012 {\em Phys. Rev. A\/}
  {\bf 85} 032318

\bibitem{darrigo2012arxiv}
D'Arrigo A, {Lo Franco} R, Benenti G, Paladino E and Falci G arXiv:1207.3294

\bibitem{coupledCT} Paladino E {\em et al}, 2010  Phys. Rev. {\bf B 81} 052502;
Paladino E {\em et al} 2011  %, A D'Arrigo, A. Mastellone and G. Falci 
New J. Phys. {\bf 13} 093037;
D'Arrigo A and Paladino E 2012 New J. Phys. {\bf 14},  053035.

\bibitem{modi2012review}
Modi K, Brodutch A, Cable H, Paterek T and Vedral V arXiv:1112.6238

\bibitem{lofrancoinprogress}
{Lo Franco} R, D'Arrigo A, Falci G, Compagno G and Paladino E \textit{in
  preparation} (2012)
      



\end{thebibliography}
\end{document}